\documentclass[epj,final]{svjour}
\usepackage{cite,latexsym}
\usepackage{psfig,epsfig}
\begin{document}
\title{Modelling Collective Opinion Formation by Means of Active Brownian
  Particles}

\author{Frank Schweitzer\inst{1,3}\thanks{\emph{Corresponding author:}
  {\tt schweitzer@gmd.de}}
  \and Janusz A. Ho{\l}yst\inst{2,3}}

\institute{GMD Institute for Autonomous intelligent Systems, Schloss
  Birlinghoven, 53754 Sankt Augustin, Germany \and Faculty of Physics,
  Warsaw University of Technology, Koszykowa 75, 00-662 Warsaw, Poland
  \and Institute of Physics, Humboldt University, Unter den Linden 6,
  10099 Berlin, Germany}

\date{Revised: January 26, 2000} 

\abstract{The concept of active
  Brownian particles is used to model a collective opinion formation
  process. It is assumed that individuals in community create a
  two-component communication field that influences the change of
  opinions of other persons and/or can induce their migration.  The
  communication field is described by a reaction-diffusion equation, the
  opinion change of the individuals is given by a master equation, while
  the migration is described by a set of Langevin equations, coupled by
  the communication field.  In the mean-field limit holding for fast
  communication we derive a critical population size, above which the
  community separates into a majority and a minority with opposite
  opinions. The existence of external support (e.g. from mass media)
  changes the ratio between minority and majority, until above a critical
  external support the supported subpopulation exists always as a
  majority. Spatial effects lead to two critical ``social'' temperatures,
  between which the community exists in a metastable state, thus
  fluctuations below a certain critical wave number may result in a
  spatial opinion separation. The range of metastability is
  particularly determined by a parameter characterizing the individual
  response to the communication field.  In our discussion, we draw
  analogies to phase transitions in physical systems.
\PACS{
{05.40.-a}{Fluctuation phenomena, random processes, noise, and Brown\-ian 
motion} \and 
{05.65.+b}{Self-organized systems} \and
{87.23.Ge}{Dynamics of social systems}
}
}
\authorrunning{Schweitzer/Ho{\l}yst} 
\titlerunning{Modelling Collective Opinion Formation ...} 
\maketitle
\newcommand{\mean}[1]{\left\langle #1 \right\rangle}
\newcommand{\abs}[1]{\left| #1 \right|}
\newcommand{\la}{\langle}
\newcommand{\ra}{\rangle}
\newcommand{\RA}{\Rightarrow}
\newcommand{\tet}{\vartheta}
\newcommand{\eps}{\varepsilon}

\newcommand{\bbox}[1]{\mbox{\boldmath $#1$}}
\newcommand{\ul}[1]{\underline{#1}}
\newcommand{\non}{\nonumber \\}
\newcommand{\no}{\nonumber}
\newcommand{\eqn}[1]{Eq. (\ref{#1})}
\newcommand{\Eqn}[1]{Eq. (\ref{#1})}
\newcommand{\eqs}[2]{eqs. (\ref{#1}), (\ref{#2})}
\newcommand{\pic}[1]{Fig. \ref{#1}}
\newcommand{\name}[1]{{\rm #1}}
\newcommand{\bib}[4]{\bibitem{#1} {\rm #2} (#4): #3.}
\newcommand{\vol}[1]{{\bf #1}}
\newcommand{\et}{{\it et al.}}
\newcommand{\fn}[1]{\footnote{ #1}}

\newcommand{\D}{\displaystyle}
\newcommand{\T}{\textstyle}
\newcommand{\SC}{\scriptstyle}
\newcommand{\SSC}{\scriptscriptstyle}

\renewcommand{\textfraction}{0.05}
\renewcommand{\topfraction}{0.95}
\renewcommand{\bottomfraction}{0.95}
\renewcommand{\floatpagefraction}{0.95}

\section{Introduction} 

In recent years, there has been a lot of interest in applications of
physical paradigms to a {\it quantitative} description of social
\cite{weidl-haag-83,galam-90,weidl-91,
  galam-91,valla-nowak-94,gilbert-doran-94,helbing-95} and economic
processes \cite{stanley96, laherrere-sornette98,
  challet-zhang98,mantegna-stanley99}

Methods of synergetics \cite{haken-78,wei-bin-91}, stochastic 
processes \cite{helbing-93,bouchaud-sornette94}, deterministic chaos 
\cite{dendrinos-sonis-90,lorenz-93,holyst-hagel-haa-weidl-96,%
  holyst-hagel-haag-97} and lattice gas models
\cite{lewenst-nowak-latane-92,kacp-holyst-96,galam97} have been
successfully applied for this purpose.
 
The formation of public opinion \cite{nowak-szam-latane-90,%
  schw-bart-pohlm-91,galam-moscovich-91,
  latane-nowak-liu-94,kacp-holyst-97,plewczynski98, kacp-holyst99} is
among the challenging problems in social science, because it reveals a
complex dynamics, which may depend on different internal and external
influences.  We mention the influence of political leaders, the biasing
effect of mass media, as well as individual features, such as persuasion
or support for other opinions.
 
A quantitative approach to the dynamics of opinion formation is given by 
the concept of \emph{social impact} \cite{lewenst-nowak-latane-92,%
  nowak-szam-latane-90}, which is based on methods similar to the
cellular automata approach \cite{wolfram-86,galam97}. The social impact
describes the force on an individual to keep or to change its current
opinion. A short outline of this model is given in Sect. 2.  The
equilibrium statistical mechanics of the social impact model was
formulated in \cite{lewenst-nowak-latane-92}, while in
\cite{kacp-holyst-96,kacp-holyst-97,kacp-holyst99} the occurrence of
phase transitions and bistability in the presence of a strong leader or
an external impact have been analysed.
 
Despite these extensive studies of the social impact model, there are 
several basic disadvantages of the concept. In particular, the social 
impact theory assumes, that the impact on an individual is 
 updated with infinite velocity, and no memory effects are 
considered. Further, there is no migration of the individuals, and any 
``spatial'' distribution of opinions refer to a ``social'', but not 
to the physical space. 
 
In fact, the model of social impact has not been developed to describe
processes of opinion diffusion and migration. In this paper, we present
an alternative approach to the social impact model of collective opinion
formation, which tries to include these features.  Our model is based on
{\em active Brownian particles}, which interact via a communication
field. This field considers the spatial distribution of the individual
opinions, further, it has a certain life time, reflecting a collective
memory effect and it can spread out in the community, modeling the
transfer of information.
 
Active Brownian particles 
\cite{lsg-mieth-rose-malch-95,lsg-schw-mieth-97,schw-agent-97} are 
Brownian particles with the ability to take up energy from the 
environment, to store it in an internal depot 
\cite{fs-eb-tilch-98-let,eb-fs-tilch-98} and to convert internal 
energy to perform different activities, such as metabolism, 
motion, change of the environment, or signal-response behavior. 
As a specific action, the active Brownian particles (or active walkers, 
within a discrete approximation) are able to generate a self-consistent 
field, which in turn influences their further movement and physical or 
chemical behavior. This non-linear feedback between the particles and the 
field generated by themselves results in an interactive structure 
formation process on the macroscopic level. Hence, these models have been 
used to simulate a broad variety of pattern formation processes in complex 
systems, ranging from physical to biological and social systems 
\cite{
lam-po-93,schwei-lsg-94,%
lam-95, 
schwei-lao-fam-97,
helb-schw-et-97,fs-98}
.
 
In Sect. 2, we specify the model of active Brownian particles for the
formation of collective opinion structures. In Sect. 3, we discuss the
limiting case of fast communication between the individuals. Further, we
investigate the influence of an external support and derive critical
parameters for the existence of subpopulations as majorities or
minorities. In Sect. 4, we investigate spatial opinion structures, and
estimate critical wave numbers for the fluctuations, which lead to a
spatial separation of the opinions. By deriving two different critical
temperatures, we draw an analogy to the theory of phase transitions.
 
\section{Stochastic Model of Opinion Change and Migration} 
 
Let us consider a 2-dimensional spatial system with the total area $A$,
where a community of $N$ individuals (members of a social group) exists.
Each of them can share one of two opposite opinions on a given subject,
denoted as $\theta_{i}=\pm1; i=1,...,N$. Here, $\theta_{i}$ is considered
as an individual parameter, representing an {\em internal degree of
  freedom}.  Within a stochastic approach, the probability
$p_{i}(\theta_{i},t)$ to find the individual $i$ with the opinion
$\theta_{i}$, changes in the course of time due to the following master
equation:
\begin{equation}  
\frac{d}{d t}p_{i}(\theta_{i},t)=\sum_{\theta_{i}'} 
w(\theta_{i}|\theta_{i}') p_{i}(\theta_{i}',t) 
- p_{i}(\theta_{i},t)\sum_{\theta_{i}'} w(\theta_{i}'|\theta_{i}). 
\label{p-eins} 
\end{equation} 
Here, $w(\theta_{i}'|\theta_{i})$ means the transition rate to change the
opinion $\theta_{i}$ into one of the possible opinions $\theta_{i}'$
during the next time step, with $w(\theta_{i}|\theta_{i})=0$. In the
considered case, there are only two possibilities, either $\theta_{i}=+1
\to \theta_{i}'=-1$, or $\theta_{i}=-1 \to
\theta_{i}'=+1$.  In the social impact theory %
\cite{lewenst-nowak-latane-92,nowak-szam-latane-90}, %
it is assumed that the change of opinions depends on the social
impact, $I_{i}$, and a ``social temperature'', $T$
\cite{kacp-holyst-96,kacp-holyst-97}. A possible ansatz for the
transition rate reads: 
\begin{equation} 
w(\theta_{i}'|\theta_{i})= \eta \, \exp\{I_{i}/T\}. 
\label{wI} 
\end{equation} 
Here, $\eta$ [1/s] defines the time scale of the transitions.  $T$
represents the erratic circumstances of the opinion change: in the limit
\mbox{$T \to 0$} the opinion change is more determined by $I_{i}$,
leading to deterministic transitions. As eq. (\ref{wI}) indicates, the
likelyhood for changing the opinion is rather small, if $I_{i}<0$. Hence, a
negative social impact on individual $i$ represents a condition for {\em
  stability}. To be specific, in the social impact theory, $I_{i}$ may
consist of three parts:
\begin{equation} 
\label{Ips} 
I_{i}=I^{p}_{i}+I^{s}_{i}+I^{ex}_{i} 
\end{equation} 
$I^{p}_{i}$ represents influences imposed on the individual by other 
members of the group, e.g. to change or to keep its opinion. $I^{s}_{i}$, 
on the other hand, is kind of a self-support for the own opinion, 
$I^{s}_{i}<0$, and $I^{ex}_{i}$ represents external influences, e.g. from 
government policy, mass media, etc. which may also support a certain 
opinion. 
 
Within a simplified approach of the social impact theory, every
individual can be ascribed a single parameter, the ``strength'', $s_{i}$.
Furthermore, a social distance $d_{ij}$ is defined, which measures the
distance between each two individuals $(i,j)$ in  a {\em
  social space} \cite{lewenst-nowak-latane-92,nowak-szam-latane-90},
which does not necessarily coincide with the physical space. It is
assumed that the impact between two individuals decreases with the social
distance in a non-linear manner. The above assumptions are included in
the following ansatz \cite{kacp-holyst-96,kacp-holyst-97}:
\begin{equation} 
I_{i}=-\theta_{i} \sum_{j=1,j\neq i}^{N} s_{j}\theta_{j}/d_{ij}^{n} \; 
-\eps s_{i}\; + e_{i}\theta_{i} 
\label{Iexpl} 
\end{equation} 
$\eps$ is the so-called self-support parameter, and $n>0$ is a model
constant. The external 
influence, $e_{i}$ may be regarded as a global preference towards one 
of the opinions. A negative social impact on individual $i$ is 
obtained, (i) if most of the opinions in its social vicinity match its 
own opinion, or (ii) if the impact resulting from opposite opinions is 
at least not large enough to compensate its self-support, or (iii) if 
the external influences do not force the individual to change its 
opinion, regardless of self-support or the impact of the community. 
 
In the form outlined above, the concept of social impact has certain
drawbacks: The social impact theory assumes that the impact on an
individual is {\em instantaneously} updated, if some opinions are changed
in the group (which basically means a communication with infinite
velocity). Spatial effects in a physical space are not considered here,
any ``spatial'' distribution of opinions refers to the social space.
Moreover, the individuals are not allowed to move. Finally, no memory
effects are considered in the social impact, the community is only
affected by the current state of the opinion distribution, regardless of
its history and past experience.
 
In this paper, we want to modify the theory by including some important
features of social systems: (i) the existence of a {\em memory}, which
reflects the past experience, (ii) an {\em exchange of information} in
the community with a {\em finite} velocity, (iii) the influence of {\em
  spatial distances} between individuals, (iv) the possibility of {\em
  spatial migration} for the individuals. It seems more realistic to us
that individuals have the chance to migrate to places where their opinion
is supported rather than change their opinion. And in most cases,
individuals are not instantaneously affected by the opinions of others,
especially if they are not in their close vicinity.
 
As a basic element of our theory, a scalar {\em spatio-temporal
  communication field} $h_{\theta}(\bbox{r},t)$ is used.
Every individual contributes permanently to this field with its opinion
$\theta_{i}$ and with its personal strength $s_{i}$
at its current spatial location $\bbox{r}_{i}$. The information generated this
way has a certain life time $1/\beta$ [s], further it can spread
throughout the system by a diffusion-like process, where $D_{h}$
[m$^{2}$/s] represents the diffusion constant for information exchange.
We have to take into account that there are two different opinions in the
system, hence the communication field should also consist of two
components, $\theta=\{-1,+1\}$, each representing one opinion.  For
simplicity, it is assumed that the information resulting from the
different opinions has the same life time and the same way of spatial
distribution; more complex cases can be considered as well.
 
The spatio-temporal change of the communication field can be summarized in 
the following equation:  
\begin{eqnarray} 
\label{hrt} 
\frac{\partial}{\partial t} h_{\theta}(\bbox{r},t) &=& 
\sum_{i=1}^{N}s_{i}\;\delta_{\theta,\theta_{i}}\;
\delta(\bbox{r}-\bbox{r}_{i})\;\nonumber\\ &&
- \;\beta h_{\theta}(\bbox{r},t) \;+ \;D_{h} \Delta h_{\theta}(\bbox{r},t). 
\end{eqnarray} 
Here, $\delta_{\theta,\theta_{i}}$ is the \name{Kronecker} Delta
indicating that the individuals contribute only to the field component
which matches their opinion $\theta_{i}$. $\delta(\bbox{r}-\bbox{r}_{i})$ means
\name{Dirac's} Delta function used for continuous variables, which
indicates that the individuals contribute to the field only at their
current position, $\bbox{r}_{i}$.
We note that this equation is a stochastic partial differential equation 
with 
\begin{equation} 
\label{n-micr} 
n^{micr}(\bbox{r},t)=\sum_{i=1}^{N} \delta(\bbox{r}-\bbox{r}_{i}) 
\end{equation} 
being the microscopic density \cite{lsg-schw-mieth-97} of the individuals
changing their position due to \eqn{langev-red}. 
Hence, the changes of the
communication field $ h_{\theta}(\bbox{r},t)$ are measured in units of a
\emph{density} of the personal strength $s_{i}$.

Instead of a social impact, the communication field $h_{\theta}(\bbox{r},t)$ 
influences the individual $i$ as follows: At a certain location 
$\bbox{r}_{i}$, the individual with opinion $\theta_{i}=+1$ is 
affected by two kinds of information: the information resulting from 
individuals who share his/her opinion, $h_{\theta=+1}(\bbox{r}_{i},t)$, and the 
information resulting from the opponents 
$h_{\theta=-1}(\bbox{r}_{i},t)$. The diffusion constant $D_{h}$ 
determines how fast he/she will receive any information, and the decay 
rate $\beta$ determines, how long a generated information will exist. 
Dependent on the {\em local} information, the individual has two 
opportunities to act: (i) it can \emph{change its opinion}, (ii) it can 
\emph{migrate} towards locations which provide a larger support of its 
current opinion. These opportunities are specified in the following.
 
For the change of opinions, we can adopt the transition probability, 
\eqn{wI}, by replacing the influence of the social impact $I_{i}$ 
with the influence of the local communication field. A possible ansatz 
reads:  
\begin{eqnarray} 
w(\theta_{i}'|\theta_{i})&=&  
\eta \, \exp\{[h_{\theta'}(\bbox{r}_{i},t)-h_{\theta}(\bbox{r}_{i},t)]/T\} \non
w(\theta_{i}|\theta_{i})&=& 0 
\label{wh} 
\end{eqnarray} 
As in \eqn{wI}, the probability to change opinion $\theta_{i}$ is rather
small, if the local field $h_{\theta}(\bbox{r}_{i},t)$, which is related to the
support of opinion $\theta_{i}$, overcomes the local influence of the
opposite opinion. This effect, however, is scaled again by the
\emph{social temperature} $T$, which is a measure for the randomness in
social interaction.  Note, that the social temperature is measured in
units of the communication field.

The movement of the individual located at space coordinate $\bbox{r}_{i}$ may
depend both on erratic circumstances and on the influence of the
communication field. Within a stochastic approach, this movement can be
described by the following overdamped Langevin equation:
\begin{equation} 
\frac{d\bbox{r}_i}{dt}=\alpha_{i} \left. 
\frac{\partial h_e(\bbox{r},t)}{\partial r}\right|_{\bbox{r}_i}  
+ \sqrt{2\,D_{n}}\;\xi(t). 
\label{langev-red} 
\end{equation}  
In the last term of \eqn{langev-red} $D_{n}$ means the spatial
diffusion coefficient of the individuals. The random influences on the
movement are modeled by a stochastic force with a $\delta$-correlated
time dependence, i.e. $\xi(t)$ is the white noise with
$\mean{\xi_{i}(t)\,\xi_{j}(t')}=\delta_{ij}\,\delta(t-t')$.

The term $h_e(\bbox{r},t)$ in \eqn{langev-red} means an {\em effective}
communication field which results from $h_{\theta}(\bbox{r},t)$ as specified
below. It follows that 
the overdamped Langevin \eqn{langev-red}  considers the response of
the individual to the {\em gradient} of the field $h_e(\bbox{r},t)$, where
$\alpha_{i}$ is the individual response parameter, weighting the
importance of the information received.  In the considered case, the
effective communication field $h_e(\bbox{r},t)$ is a certain function of both
components, $h_{\pm 1}(\bbox{r},t)$, of the communication field, see \eqn{hrt}. One
can consider different types of response, for example the following:
\begin{enumerate} 
\item[(i)] The individuals try to move towards locations which provide 
  the most support for their current opinion $\theta_{i}$. In this 
  case, they only count on the information which matches their opinion, 
  $h_e(\bbox{r},t)= h_{\theta}(\bbox{r},t)$, and follow the local ascent of the field 
  $(\alpha_{i} > 0)$. 
\item[(ii)] The individuals try to move away from locations which provide 
  any negative pressure on their current opinion $\theta_{i}$.  In this 
  case, they count on the information resulting from opposite opinions 
  $(\theta')$, $h_e(\bbox{r},t)= h_{\theta'}(\bbox{r},t)$, and follow the local 
  descent of the field $(\alpha_{i} < 0)$. 
\item[(iii)] The individuals try to move away from locations, if they are 
  forced to change their current opinion $\theta_{i}$, but they can 
  accept a vicinity of opposite opinions, as long as these are not 
  dominating. In this case, they count on the information resulting from 
  both supporting and opposite opinions, and the local difference between 
  them is important: $h_e(\bbox{r},t)= [h_{\theta}(\bbox{r},t)-h_{\theta'}(\bbox{r},t)]$ 
  with $\alpha_{i} > 0$. 
\end{enumerate} 
Additionally, the response parameter can also consider that the response
occurs only, if the absolute value of the effective field is locally above
a certain threshold $h_{thr}$: \mbox{%
$\alpha_{i}=\Theta\left[\abs{h_e(\bbox{r},t)}-h_{thr}\right]$}, with $\Theta[y]$
being the Heavyside function: $\Theta=1$, if $y>0$, otherwise $\Theta=0$.
We note that for the further discussions in Sect. 3 and 4, we assume
$h_e(\bbox{r},t)= h_{\theta}(\bbox{r},t)$ for the effective communication field (case
i), while $\alpha_{i}\equiv\alpha$ is treated as a positive constant
independent of $i$ and $h_e(\bbox{r},t)$.

In order to summarize our model, we note the non-linear feedback between
the individuals and the communication field as shown in \pic{caus4a}. The
individuals generate the field, which in turn influences their further
movement and their opinion change. In terms of synergetics, the field
plays the role of an order parameter, which couples the individual
actions, and this way initiates spatial structures and coherent behavior
within the social group.  
\begin{figure}[ht]
\centerline{\psfig{figure=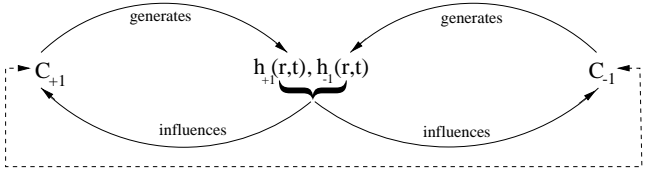,width=8.5cm}}
\caption[]{Circular causation between the individuals with different
  opinions, $C_{-1}$, $C_{+1}$ and the two-component communication field,
  $h_{\theta}(\bbox{r},t)$. \label{caus4a}}
\end{figure}

The complete dynamics of the community can be formulated in terms of the 
canonical $N$-particle distribution function  
\begin{equation} 
\label{prt} 
P(\ul{\theta},\ul{r},t)=P(\theta_{1},\bbox{r}_{1},...,\theta_{N},\bbox{r}_{N},t), 
\end{equation} 
which gives the probability to find the $N$ individuals with the 
opinions $\theta_{1},...,\theta_{N}$ in the vicinity of $\bbox{r}_1,....,\bbox{r}_N$ 
on the surface $A$ at time $t$.  Considering both opinion changes and 
movement of the individuals, the master equation for 
$P(\ul{\theta},\ul{r},t)$ reads: 
\begin{eqnarray} 
\frac{\D \partial}{\D \partial t}P(\ul{\theta},\ul{r},t) =   
\sum_{\ul{\theta'} \neq \ul{\theta}} \Big[
w(\ul{\theta}|\ul{\theta}') P(\ul{\theta}',\ul{r},t) -  
w(\ul{\theta}'|\ul{\theta}) P(\ul{\theta},\ul{r},t) \Big] \nonumber\\  
- \sum_{i=1}^N\Big[\nabla_i\,\left( \alpha \, 
\nabla_i h_{\theta}(\bbox{r},t) \, 
P(\ul{\theta},\ul{r},t)\right) - D_n\,\Delta_i P(\ul{r},\ul{\theta},t)
\Big]  \nonumber\\  
\label{master} 
\end{eqnarray} 
The first line of the right-hand side of \eqn{master} describes the 
``gain'' and ``loss'' of individuals (with the coordinates 
$\bbox{r}_1,...,\bbox{r}_N$) due to opinion changes, where 
$w(\ul{\theta}|\ul{\theta}')$ means any possible transition within the 
opinion distribution $\ul{\theta}'$ which leads to the assumed 
distribution $\ul{\theta}$. The second line describes the change of 
the probability density due to the motion of the individuals on the 
surface. \Eqn{master} together with \eqs{hrt}{wh} forms a complete
description of our system.

\section{The Case of Fast Communication} 
\subsection{Derivation of Mean Value Equations} 
Let us first restrict to the case of very fast exchange of information 
in the system. Then, spatial inhomogenities are equalized immediately, 
hence, the communication field $h_{\theta}(\bbox{r},t)$ can be approximated by a 
mean field $\bar{h}_{\theta}(t)$: 
\begin{equation} 
  \label{hat} 
  \bar{h}_{\theta}(t)=\frac{1}{A}\int\limits_{A} h_{\theta}(\bbox{r},t)\; dr^{2}, 
\end{equation} 
where $A$ means the system size. The equation for the mean field 
$\bar{h}_{\theta}(t)$ results from eq. (\ref{hrt}): 
\begin{equation} 
\frac{\partial \bar{h}_{\theta}(t)}{\partial t} = 
- \beta\bar{h}_{\theta}(t) + s \bar{n}_{\theta} 
\label{hat-t} 
\end{equation} 
with $s_{i}\equiv s$ and the mean density 
\begin{equation} 
  \label{bar-ntheta} 
\bar{n}_{\theta}=\frac{N_{\theta}}{A}\;\;;\;\;\; 
\bar{n}=\frac{N}{A},
\end{equation}
where the number of individuals with a given opinion $\theta$ fulfils the
condition 
\begin{equation}
\label{sum}
\sum_{\theta}N_{\theta}=N_{+1}+N_{-1}=N={\rm const.}   
\end{equation} 
We note that in the mean--field approximation no spatial gradients in 
the communication field exist. Hence, there is no additional driving 
force for the individuals to move, as assumed in 
\eqn{langev-red}. Such a situation can be 
imagined for communities existing in very small systems with small 
distances between different groups. In particular, in such small 
communities also the assumption of a fast information exchange 
holds. Thus, in this section, we restrict our discussion to 
subpopulations with a certain opinion rather than to individuals at 
particular locations. 
 
Let $p(N_{\theta},t)$ denote the probability to find $N_{\theta}$ 
individuals in the community which shares opinion $\theta$.
The master equation for $p(N_{+1},t)$ explicitely reads: 
\begin{eqnarray} 
\frac{\partial}{\partial t}p(N_{+1},t) & = &  
W(N_{+1}|N_{+1}-1)\, p(N_{+1}-1,t) \nonumber \\ 
& & + W(N_{+1}|N_{+1}+1) \, p(N_{+1}+1,t) \nonumber \\ 
& & - p(N_{+1},t)\left[W(N_{+1}+1|N_{+1}) \right. \nonumber \\ 
& & + \left. W(N_{+1}-1|N_{+1})\right]. 
\label{master-pN} 
\end{eqnarray} 
The transition rates $W(M|N)$ appearing in \eqn{master-pN} are assumed to
be proportional to the probability to change a given opinion, eq.
(\ref{wh}), and to the number of individuals which can change their
opinion into the given direction:
\begin{eqnarray} 
W(N_{+1}+1|N_{+1})& = & N_{-1} \,  
\eta \, \exp{\{(\bar{h}_{+1}-\bar{h}_{-1})/T\}}, \\ 
W(N_{+1}-1|N_{+1})& = & N_{+1}\,\eta \, 
\exp{\{-(\bar{h}_{+1}-\bar{h}_{-1})/T\}}. \nonumber
\label{wN} 
\end{eqnarray} 
The mean values for the number of individuals with a certain 
opinion can be derived from the master equation (\ref{master-pN}) 
\begin{equation} 
\langle N_{\theta}(t) \rangle =  
\sum_{\{N_{\theta}\}} N_{\theta}\; p(N_{\theta},t), 
\label{np} 
\end{equation} 
where the summation is over all possible numbers of $N_{\theta}$ which
obey the condition eq. (\ref{sum}). From eq. (\ref{np}), the
deterministic equation for the change of $\mean{N_{\theta}}$ can be
derived in the first approximation as follows \cite{schw-lsg-eb-ulb} (see
also \cite{weidl-haag-83,weidl-91,helbing-95}):
\begin{equation} 
\frac{d}{dt}\left\langle N_{\theta} \right\rangle =  
\mean{W(N_{\theta}+1|N_{\theta})  
- W(N_{\theta} -1|N_{\theta})}
\label{ntheta} 
\end{equation} 
For $N_{+1}$, this equation reads explicitely: 
\begin{eqnarray} 
\frac{d}{dt}\left\langle N_{+1} \right\rangle& =& \left\langle 
N_{-1} \,\eta \, \exp \left[\frac{\bar{h}_{+1}(t) 
-\bar{h}_{-1}(t)}{T} \right] \right. \nonumber \\  
& & \left.  - N_{+1} \,\eta \, \exp \left[- \frac{\bar{h}_{+1}(t) 
-\bar{h}_{-1}(t)}{T} \right] \right\rangle 
\label{n1det} 
\end{eqnarray} 
Introducing now the fraction of a {\em subpopulation} with opinion
$\theta$, $x_{\theta}=\langle N_{\theta}\rangle/N$, and using the standard
approximation to factorize eq. (\ref{n1det}), we can write it as:
\begin{eqnarray} 
\label{xdet} 
\dot{x}_{+1} &=&(1-x_{+1})\,\eta \, \exp(a) - x_{+1}\,\eta \, \exp(-a),\\  
a&=& \left[\bar{h}_{+1}(t)-\bar{h}_{-1}(t)\right]/ T. \nonumber 
\end{eqnarray} 
Via $\Delta \bar{h}(t)= \bar{h}_{+1}-\bar{h}_{-1}$, this equation
is coupled with the equation 
\begin{equation} 
\label{hx} 
\Delta \dot{\bar{h}} = -\beta\, \Delta\bar{h} + s \bar{n} \Big(2 x_{+1}
-1\Big) 
\end{equation} 
which results from eq. (\ref{hat-t}) for the two field components.

\subsection{Critical and Stable Subpopulation Sizes} 
Within a quasistationary approximation, we can assume that the 
communication field \emph{relaxes faster} than the distribution of the 
opinions into a stationary state. Hence, with 
$\dot{\bar{h}}_{\theta}=0$, we find from \eqn{hat-t}: 
\begin{equation} 
\label{ha} 
\begin{array}{lcl} 
\bar{h}_{+1}^{stat}=\frac{\D  s\,\bar{n}}{\D \beta}x_{+1} &;&  
\bar{h}_{-1}^{stat}=\frac{\D s\,\bar{n}}{\D \beta}(1-x_{+1})\\ 
a=\kappa \left(x_{+1}-\frac{\D 1}{\D 2}\right) &\mbox{with}&
\kappa=\frac{\D 2s\,\bar{n}}{\D \beta T} 
\end{array} 
\end{equation} 
Here, the parameter $\kappa$ includes the specific \emph{internal
  conditions} within the community, such as the total population size,
the social temperature, the individual strength of the opinions, or
the life time of the information generated. Inserting $a$ from eq.
(\ref{ha}) into eq. (22), a closed equation for $\dot{x}_{\theta}$ is
obtained, which can be integrated with respect to time (\pic{crit}a). We
find that, depending on $\kappa$, different stationary values for the
fraction of the subpopulations exist. For the critical value, $\kappa^{c}
=2$, the stationary state can be reached only asymptotically.
\pic{crit}(b) shows the stationary solutions, $\dot{x}_{\theta}=0$,
resulting from the equation for $x_{+1}$:
\begin{equation} 
  \label{xkrit} 
(1-x_{+1})\,\exp\left[\kappa\, x_{+1}\right]=  
  x_{+1}\,\exp\left[\kappa\,(1-x_{+1})\right] 
\end{equation} 
\begin{figure}[ht] 
\centerline{ 
\psfig{figure=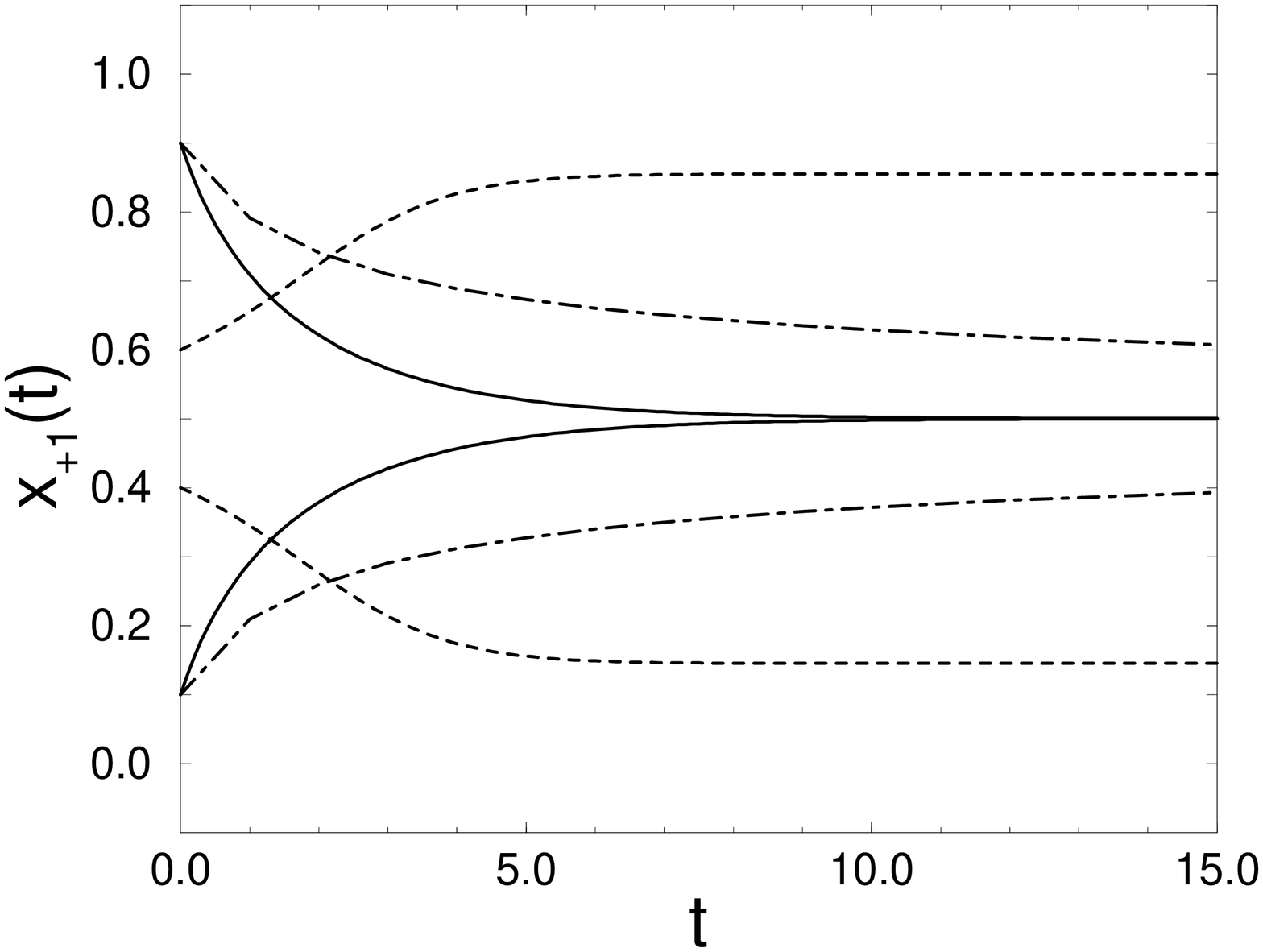,height=7.0cm,angle=-90}} 
\centerline{ 
\psfig{figure=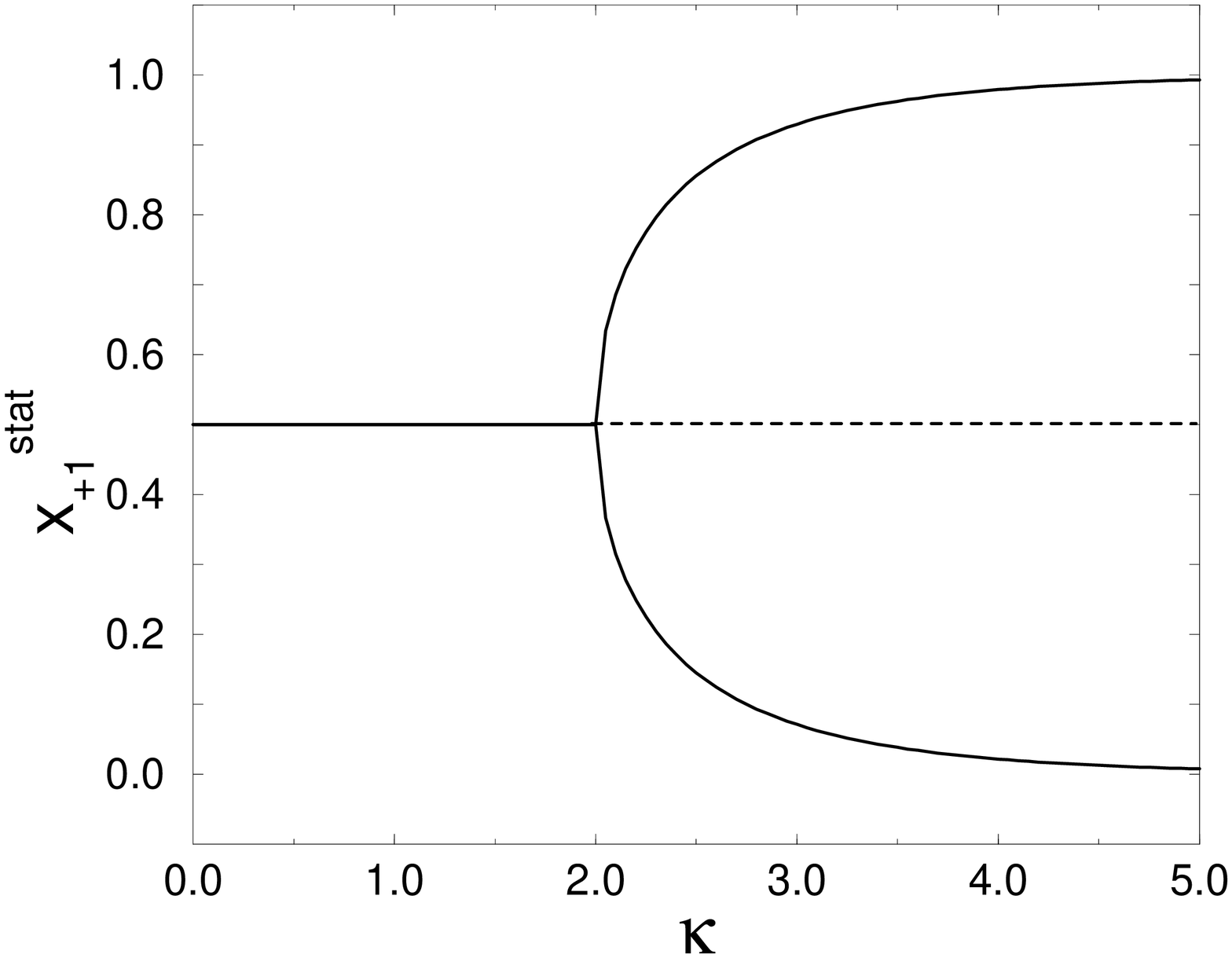,height=7.0cm,angle=-90}} 
\caption[fig1]{ 
  (a: top) Time dependence of the fraction $x_{+1}(t)$ (eq. \ref{xdet}) 
  of the subpopulation with opinion $+1$ for different initial conditions
  and for three different values of 
  $\kappa$: 1.0 (solid line); 2.0 (dot-dashed line), 3.0 (dashed line). 
  (b: bottom) Stationary solutions for $x_{+1}$ (eq. \ref{xkrit}) for 
  different values of $\kappa$. The bifurcation at the critical value 
  $\kappa^{c}=2$ is clearly visible.} 
  \label{crit} 
\end{figure} 
 
For $\kappa < 2$, $x_{+1}=0.5$ is the only stationary solution, which
means a stable community where both opposite opinions have the same
influence. However, for $\kappa>2$, the equal distribution of opinions
becomes unstable, and a separation process towards a preferred opinion is
obtained, where $x_{\pm1}=0.5$ plays the role of a separation line. We
find now two stable solutions where both opinions coexist with different
shares in the community, as shown in \pic{crit}. Hence, each
subpopulation can exist either as a
\emph{majority} or as a \emph{minority} within the community. Which of
these two possible situations is realized, depends in a deterministic
approach on the initial fraction of the subpopulation. For initial values
of $x_{+1}$ below the separatrix, $0.5$, the minority status will be most
likely the stable situation, as \pic{crit}(a) shows.
 
The bifurcation occurs at $\kappa^{c} =2$, where the former stable 
solution $x_{+1}=0.5$ becomes unstable. From the condition $\kappa =2$ we 
can derive a \emph{critical population size}, 
\begin{equation} 
  \label{n-crit} 
 N^{c}= \beta \,A\,T/s,  
\end{equation} 
where for larger populations an equal fraction of opposite opinions is 
certainly unstable. If we consider e.g. a \emph{growing community} with 
fast communication, then both contradicting opinions are balanced, as 
long as the population number is small. However, for $N>N^{c}$, i.e. 
after a certain population growth, the community tends towards one of 
these opinions, thus necessarily separating into a majority and a 
minority.  Which of these opinions would be dominating, depends on small 
fluctuations in the bifurcation point, and has to be investigated within 
a stochastic approach. We note that eq.  (\ref{n-crit}) for the critical 
population size can be also interpreted in terms of a critical social 
temperature, which leads to an opinion separation in the community. This 
will be discussed in more detail in Sect. 4. 
 
From \pic{crit}(b), we see further, that the stable coexistence between 
majority and minority breaks down at a certain value of $\kappa$, where 
almost the whole community shares the same opinion. From \eqn{xkrit} it 
is easy to find that e.g.  $\kappa \approx 4.7$ yields 
$x_{+1}\approx \{0.01;0.99\}$, which means that about $99\%$ of the 
community share either opinion $+1$ or $-1$. 
    
\subsection{Influence of External Support} 
Now, we discuss the situation that the symmetry between the two
opinions is broken due to external influences on the individuals. We may
consider two similar cases: (i) the existence of a {\em strong leader} in
the community, who possesses a strength $s_{l}$ which is much larger than
the usual strength $s$ of the other individuals, (ii) the existence of an
external field, which may result from government policy, mass media, etc.
which support a certain opinion with a strength $s_{m}$.

The additional influence
\mbox{$s_{ext}:=\{s_{l}/A,s_{m}/A\}$} mainly effects the communication
field, eq.  (\ref{hrt}), due to an extra contribution, normalized by the
system size $A$. 

If we assume an external support of opinion $\theta=+1$,
the corresponding field equation in the mean--field limit (eq. \ref{hat-t})
and the stationary solution (eq.  \ref{ha}) are changed as follows:
\begin{eqnarray} 
  \label{h1ch} 
\dot{\bar{h}}_{+1}&=& -\beta\bar{h}_{+1}(t)+s\,\bar{n}\,x_{+1}+s_{ext}
\non
\bar{h}_{+1}^{stat}&=&\frac{\D s\,\bar{n}}{\D \beta}x_{+1}+ 
\frac{\D s_{ext}}{\D \beta} \\ 
a&=&\kappa\left(x_{+1}-\frac{\D 1}{\D 2}\right) 
+\frac{\D s_{ext}}{\D \beta T}. \nonumber
\end{eqnarray} 
Hence, in \eqn{xkrit} which determines the stationary solutions, the 
arguments are shifted by a certain value: 
\begin{eqnarray} 
  \label{xkrit-ex} 
(1-x_{+1})\,\exp\left[\kappa\,x_{+1}+\frac{s_{ext}}{\beta T}\right]= \non  
  x_{+1}\,\exp\left[\kappa\,(1-x_{+1})-\frac{s_{ext}}{\beta T}\right] 
\end{eqnarray} 
\pic{crit-ex} shows how the critical and stable subpopulation sizes 
change for subcritical and supercritical values of $\kappa$, dependent on 
the strength of the external support. 
\begin{figure}[htbp] 
\centerline{ 
\psfig{figure=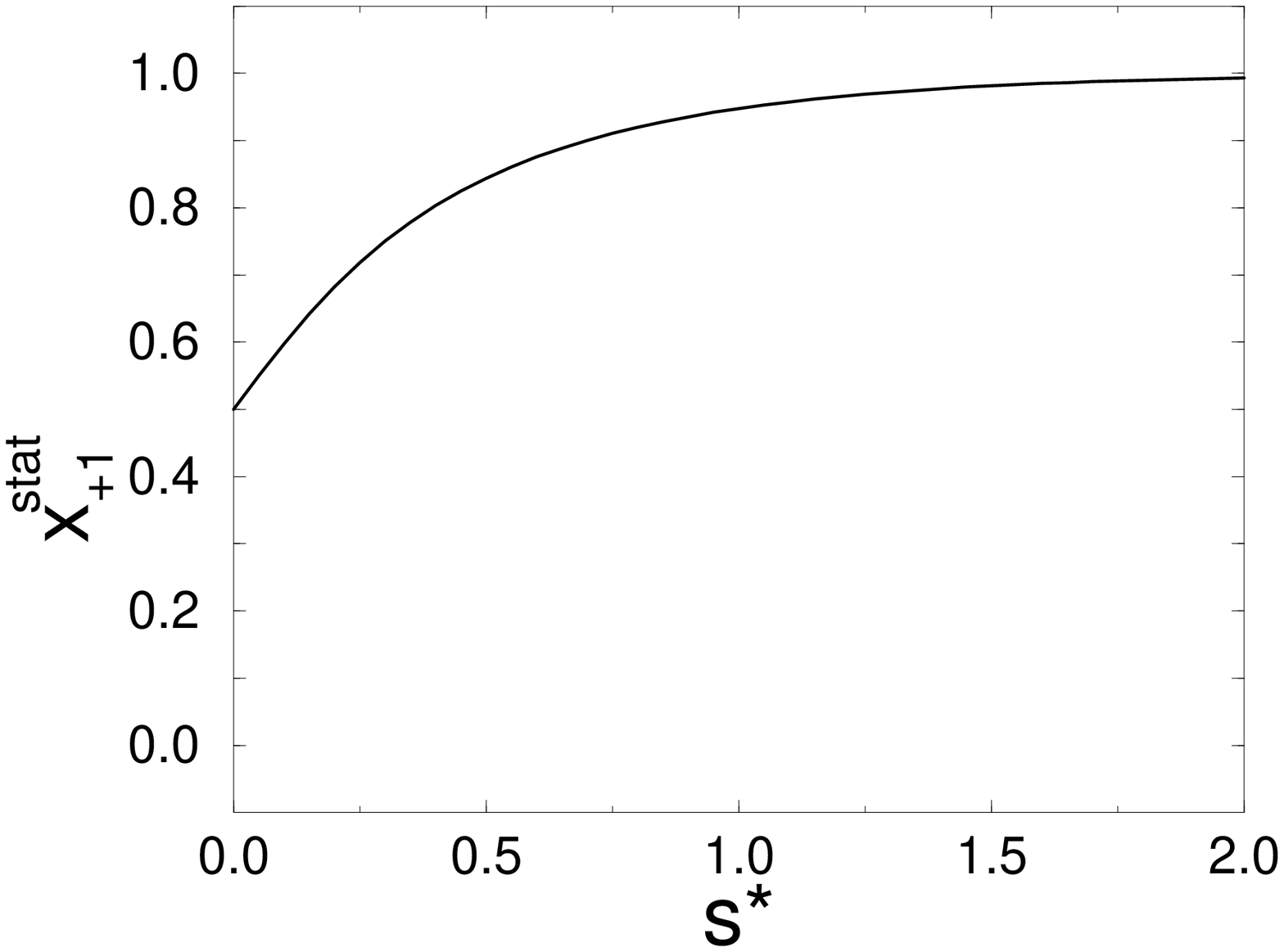,width=7.cm}}
\centerline{
\psfig{figure=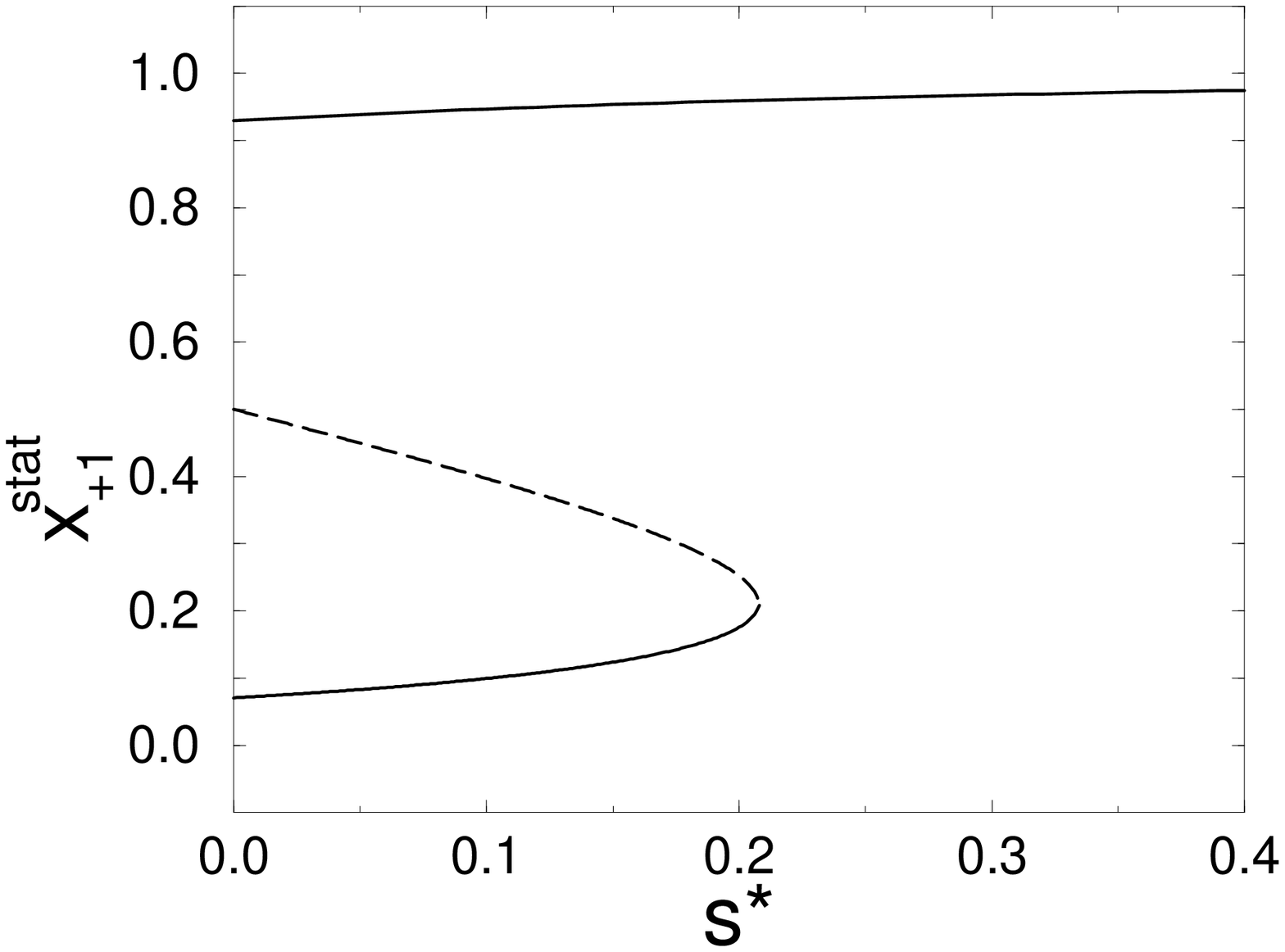,width=7.cm}} 
\caption[fig2]{ 
  Stable fraction of the subpopulation, $x_{+1}^{stat}$, as a function of
  the strength $s^{\star}=s_{ext}/\beta T$ of the external support.  (a:
  top) $\kappa =1$, (b: bottom) $\kappa=3$. The dashed line in (b)
  represents the separation line for the initial conditions, which lead
  either to a minority or a majority status of the subpopulation.}
\label{crit-ex} 
\end{figure} 
 
For $\kappa <\kappa^{c}$ (\pic{crit-ex} a), we see that there is still 
only one stable solution, but with an increasing value of $s_{ext}$, the 
supported subpopulation exists as a majority. For $\kappa > 
\kappa^{c}$ (\pic{crit-ex} b), we observe again two possible stable 
situations for the supported subpopulation, either a minority or a 
majority status. But, compared to \pic{crit}(b), the symmetry between 
these possibilities is now broken due to the external support, which 
increases the region of initial conditions leading to a majority status. 
 
Interestingly, at a critical value of $s_{ext}$, the possibility of a 
minority status completely vanishes. Hence, for a certain supercritical 
external support, the supported subpopulation will grow towards a 
majority, regardless of its initial population size, with no chance for 
the opposite opinion to be established. This situation is quite often 
realized in communities with one strong political or religious leader 
(``fundamentalistic dictatorships''), or in communities driven by 
external forces, such as financial or military power (``banana 
republics''). 
 
The value of the critical external support, $s_{ext}^{c}$, of course 
depends on $\kappa$, which summarizes the internal situation in terms of 
the social temperature, or the population size, etc. From \eqn{xkrit-ex} 
we can derive the condition for which two of the three possible solutions 
coincide, thus determining the relation between $s_{ext}^{c}$ and 
$\kappa$ as follows:  
\begin{equation} 
\label{s-crit} 
s^{\star}_{c}=\frac{s_{ext}^{c}}{\beta T}=\frac{1}{2}
\ln \left[ \frac{1-\sqrt{1-\frac{\D 2}{\D \kappa}}}{1 
+\sqrt{1-\frac{\D 2}{\D \kappa}}} \right] 
+ \frac{1}{2} \kappa \sqrt{1-\frac{\D 2}{\D \kappa}} 
\end{equation} 
\pic{s-crit-k} shows how much external support is needed to paralyze a
community with a given internal situation ($\kappa$) by one ruling
opinion. As one can see, the critical external support is an increasing
function of the parameter $\kappa$, meaning that it is more difficult
to paralyse a society with strong interpersonal interactions.
\begin{figure}[htbp] 
\centerline{ 
\psfig{figure=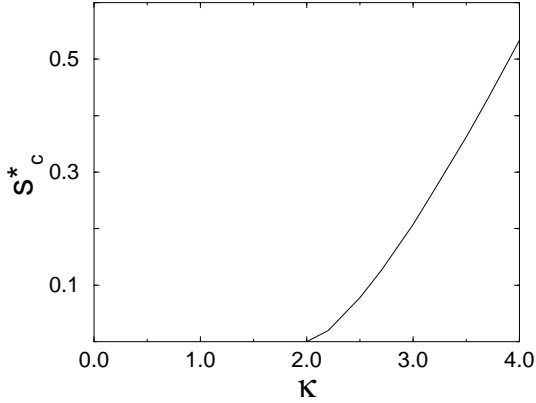,width=7.cm}} 
\caption[fig3]{ 
  Critical external support $s^{\star}_{c}$ (eq. \ref{s-crit}) as a
  function of $\kappa$.}
\label{s-crit-k} 
\end{figure} 

Let us conclude the discussion of the phase transition in the mean field
limit, presented in this section. With respect to the social impact
theory \cite{nowak-szam-latane-90,valla-nowak-94,
  lewenst-nowak-latane-92,kacp-holyst-96,kacp-holyst99}, we note that
phase transitions have been not considered there, since the focus was on
other phenomena so far.  On the other hand, for the case of two opinions,
our results well correspond to those obtained by Weidlich and Haag in a
model of collective opinion formation \cite{weidl-haag-83,weidl-91}.
There, a master equation and appropriate {\it utility potentials} are
used to analyse the mean-field dynamics of interacting populations.
Similar to our model, the approach used in \cite{weidl-haag-83,weidl-91}
leads to a phase transition and a corresponding bifuraction diagram. The
main difference between both models is in the interpretation of the
bifurcation parameter $\kappa$.  In our case, $\kappa$ results from other
model parameters, e.g. from the mean ``social strength'' $s$ which plays a
role of a coupling constant between the opinions and the communication
field.  In the model of Weidlich and Haag, on the other hand, this
parameter is interpreted as a derivative of {\it utility potentials}. Our
analytic result for the critical external support, \eqn{s-crit}, is in
qualitative agreement with the stability analysis and the computer
simulations presented in \cite{weidl-haag-83,weidl-91}.
 
\section{Critical Conditions for Spatial Opinion Separation} 
In the previous section, the existence of critical parameters, such as
$\kappa^{c}$ or $s_{ext}^{c}$, has been proved for a community with fast
communication, where no inhomogenities in the communication field can
exist. In the more realistic case, however, we have finite diffusion
coefficients for the information, and the mean-field approximation, eq.
(\ref{hat-t}), is no longer valid. Instead of focussing on the
subpopulation sizes, we now need to consider the \emph{spatial
  distribution} of individuals with opposite opinions.
 
Starting with the canonical N--particle distribution function,
$P(\ul{\theta},\ul{r},t)$, eq. (\ref{master}), the spatio-temporal
density of individuals with opinion $\theta$ can be obtained as follows:
\begin{eqnarray} 
n_{\theta}(\bbox{r},t)&=&\int \sum_{i=1}^N \delta_{\theta,\theta_{i}}\; 
\delta(\bbox{r}-\bbox{r}_{i})\times \nonumber \\
&& \times P(\theta_{1},\bbox{r}_{1}...,\theta_{N},\bbox{r}_{N},t)\; 
d\bbox{r}_1 ... d\bbox{r}_{N} 
\label{dens} 
\end{eqnarray} 
Integrating eq. (\ref{master}) according to eq. (\ref{dens}) and
neglecting higher order correlations, we obtain the following
reaction-diffusion equation for $n_{\theta}(\bbox{r},t)$
\begin{eqnarray} 
\label{fpe-t} 
\frac{\partial}{\partial t}\,n_{\theta}(\bbox{r},t) = &  
- & \nabla\,\Big[n_{\theta}(\bbox{r},t)\, \alpha \,
\nabla h_{\theta}(\bbox{r},t)\Big]  
+ D_n\,\Delta n_{\theta}(\bbox{r},t) \nonumber \\ 
& - & \sum_{\ul{\theta'} \neq \ul{\theta}} \Big[ 
w(\theta'|\theta)\,n_{\theta}(\bbox{r},t)  
+ w(\theta|\theta')\,n_{\theta'}(\bbox{r},t)\Big] \non 
\end{eqnarray}
with the transition rates obtained from eq. (\ref{wh}): 
\begin{eqnarray} 
w(\theta'|\theta)&=&  
\eta \, \exp\{[h_{\theta'}(\bbox{r},t)-h_{\theta}(\bbox{r},t)]/T\} \\
w(\theta|\theta)&=&0 \nonumber 
\label{wgr} 
\end{eqnarray} 
With $\theta=\{+1,-1\}$, Eq. (\ref{fpe-t}) is a set of two
reaction-diffusion equations, coupled both via $n_{\theta}(\bbox{r},t)$
and $h_{\theta}(\bbox{r},t)$.  Inserting the densities
$n_{\theta}(\bbox{r},t)$ and neglecting any external support, eq.
(\ref{hrt}) for the spatial communication field can be transformed into
the linear deterministic equation:
\begin{equation} 
\frac{\partial}{\partial t} h_{\theta}(\bbox{r},t)= 
s\, n_{\theta}(\bbox{r},t)\; 
- \;\beta\, h_{\theta}(\bbox{r},t) \;+ \; 
D_{h} \Delta h_{\theta}(\bbox{r},t) 
\label{hrt-m} 
\end{equation} 
The solutions for the spatio-temporal distributions of individuals and
opinions are now determined by the four coupled equations, \eqn{fpe-t}
and \eqn{hrt-m}. For our further discussion, we assume again that the
spatio-temporal communication field \emph{relaxes faster} than the
related distribution of individuals into a quasi-stationary equilibrium.
The field $h_{\theta}(\bbox{r},t)$ should still depend on time and space
coordinates, but, due to the fast relaxation, there is a fixed relation
to the spatio-temporal distribution of individuals. Further, we neglect
the independent diffusion of information, assuming that the spreading of
opinions is due to the migration of the individuals. From \eqn{hrt-m}, we
find with $\dot{h}_{\theta}(\bbox{r},t)=0$ and $D_{h}=0$:
\begin{equation} 
  \label{fixed} 
  h_{\theta}(\bbox{r},t)=\frac{s}{\beta}\,n_{\theta}(\bbox{r},t) 
\end{equation} 
which can now be inserted into \eqn{fpe-t}, thus reducing the set of 
coupled equations to two equations.  
 
The homogeneous solution for $n_{\theta}(\bbox{r},t)$ is given by the mean 
densities: 
\begin{equation} 
  \label{hom} 
\bar{n}_{\theta}=\mean{n_{\theta}(\bbox{r},t)}=\frac{\bar{n}}{2}
\end{equation} 
Under certain conditions however, the homogeneous state becomes unstable
and a spatial separation of opinions occurs. In order to investigate
these critical conditions, we allow small fluctuations around the
homogeneous state $\bar{n}_{\theta}$:
\begin{equation} 
\label{small-f} 
n_{\theta}(\bbox{r},t)=\bar{n}_{\theta}+ \delta n_{\theta}\;;\quad  
\left| \frac{\delta n_{\theta}}{\bar{n}_{\theta}} \right| \ll 1 
\end{equation} 
Inserting \eqn{small-f} into \eqn{fpe-t}, a linearization gives: 
\begin{equation} 
\label{linear} 
\frac{\partial \delta n_{\theta}}{\partial t}=\left[ D_{n} -  
\frac{\alpha s\,\bar{n}}{2 \beta}\right]\;  
\Delta \delta n_{\theta} + 
\left[\frac{\eta \,s\,\bar{n}}{\beta T}-\eta \right]\;  
\, (\delta n_{\theta} -\delta n_{-\theta})
\end{equation} 
With the ansatz 
\begin{equation} 
\label{ansatz} 
\delta n_{\theta} \sim \exp \left( 
\lambda t+i \bbox{k} \bbox{r} \right) 
\end{equation} 
we find from \eqn{linear} the dispersion relation $\lambda(\bbox{k})$ for
small inhomogeneous fluctuations with wave vector $\bbox{k}$. This
relation yields two solutions:
\begin{equation} 
\label{dispers} 
\begin{array}{l} 
\lambda_{1}(\bbox{k})= - k^{2}\,C+2B\;;\;\; 
\lambda_{2}(\bbox{k})= - k^{2}\,C \\  
B=\frac{\D \eta \,s\,\bar{n}}{\D \beta T}-\eta \;;\;\; 
C=D_{n} - \frac{\D \alpha s\,\bar{n}}{\D 2 \beta} 
\end{array} 
\end{equation} 
For homogeneous fluctuations we obtain from \eqn{dispers} 
\begin{equation} 
\label{l-hom}
\lambda_1= \frac{2\,\eta\, s \,\bar{n}}{\beta T}-2\,\eta \mbox{;}  
\quad \lambda_2=0 \quad \mbox{for}\; \bbox{k}=0 
\end{equation} 
which means that the homogeneous system is marginally stable as long as 
$\lambda_1 <0$, or $s\,\bar{n}/\beta T<1$. This result agrees with 
the condition $\kappa <2$ obtained from the previous mean field 
investigations in Sect. 3. The condition $\kappa=2$ or $B=0$, 
respectively, defines a \emph{critical social temperature} 
\begin{equation} 
  \label{crit-temp} 
  T^{c}_{1}= \frac{s\,\bar{n}}{\beta} 
\end{equation} 
For temperatures $T<T^{c}_{1}$, the homogeneous state
\mbox{$n_{\theta}(\bbox{r},t)= \bar{n}/2$}, where individuals of both
opinions are equally distributed, becomes unstable and the spatial
separation process occurs.  This is in direct analogy to the phase
transition obtained from the Ising model of a ferromagnet. Here, the
state with $\kappa < 2$ or $T>T^{c}_{1} $, respectively, corresponds to
the \emph{paramagnetic} or disordered phase, while the state with $\kappa
> 2$ or $T<T^{c}_{1} $, respectively, corresponds to \emph{ferromagnetic}
ordered phase.

The conditions of \eqn{l-hom} denote a \emph{homogeneous} stability
condition.  To obtain stability against inhomogeneous fluctuations of
wave vector $\bbox{k}$, the two conditions $\lambda_1(\bbox{k}) \leq 0$
and $\lambda_2(\bbox{k}) \leq 0$ have to be satisfied.

Taking into account the
critical temperature $T^{c}_{1}$, \eqn{crit-temp}, we can rewrite these
conditions, \eqn{dispers}, as follows:
\begin{eqnarray} 
\label{t1-D} 
\bbox{k}^2\,\Big(D_{n}-D_{n}^{c}\Big) 
-2\,\eta\,\left(\frac{T^{c}_{1}}{T}-1\right) &\geq& 0 \non 
\bbox{k}^2\,\Big(D_{n}-D_{n}^{c}\Big) &\geq& 0 
\end{eqnarray} 
Here, a \emph{critical diffusion coefficient} $D_{n}^{c}$ for the
individuals appears, which results 
from the condition $C=0$: 
\begin{equation}  
\label{Dc} 
D^{c}_{n}=\frac{\alpha}{2}\; \frac{s\,\bar{n}}{\beta} 
\end{equation}
Hence, the condition
\begin{equation}
  \label{diff-crit}
  D_{n}>D^{c}_{n}
\end{equation}
denotes a second stability condition. In order to explain its meaning,
let us consider that the diffusion coefficient of the individuals,
$D_{n}$, may be a function of the social temperature, $T$. This sounds
reasonable since the social temperature is a measure of randomness in
social interaction, and an increase of such a randomness 
lead to an increase of a random spatial migration. The simplest relation
for a function $D_{n}(T)$ is the linear one, $D_{n}=\mu T$. By assuming
this, we may rewrite \eqn{t1-D} using a \emph{second critical
  temperature}, $T_{2}^{c}$ instead of a critical diffusion coefficient
$D_{n}^{c}$:
\begin{eqnarray} 
\label{t1-t2} 
\bbox{k}^2\,\mu\,\Big(T-T^{c}_{2}\Big) 
-2\,\eta\,\left(\frac{T^{c}_{1}}{T}-1\right) &\geq& 0 \non
\bbox{k}^2\,\mu\,\Big(T-T^{c}_{2}\Big) &\geq& 0 
\end{eqnarray} 
The second critical temperature $T^{c}_{2}$ reads as follows:
\begin{equation}  
\label{temp-2} 
T^{c}_{2}=\frac{\alpha}{2\mu}\; \frac{s\,\bar{n}}{\beta} 
=\frac{\alpha}{2\mu}\;T^{c}_{1} 
\end{equation}
The occurence of two critical social temperatures $T_{1}^{c}$,
$T_{2}^{c}$ allows a more detailed discussion of the stability
conditions. Therefore, we have to consider two separate cases of
\eqn{temp-2}: (1) \mbox{$T^{c}_{1} > T^{c}_{2}$} and (2) \mbox{$T^{c}_{1}
  < T^{c}_{2}$}, which correspond either to the condition $\alpha <
2\mu$, or $\alpha > 2\mu$, respectively.

In the first case, \mbox{$T^{c}_{1} > T^{c}_{2}$}, we can discuss three
ranges of the temperature $T$:
\begin{enumerate} 
\item[(i)] For $T > T^{c}_{1}$ both eigenvalues $\lambda_1(\bbox{k})$ and
  $\lambda_2(\bbox{k})$, \eqn{dispers}, are nonpositive for all wave vectors
  $\bbox{k}$ and the homogenous solution $\bar{n}/2$ is {\it completely
    stable}.
\item[(ii)] For $ T^{c}_{1} > T > T^{c}_{2}$ the eigenvalue 
  $\lambda_2(\bbox{k})$ is still nonpositive for all values of $\bbox{k}$,  
  but the eigenvalue $\lambda_1(\bbox{k})$ is negative only for 
  wave vectors that are larger than some critical value $\bbox{k}^2 > 
  \bbox{k}_{c}^2$: 
 \begin{equation} 
\label{k-crit} 
  \bbox{k}_{c}^2 = \frac{2\,\eta}{\mu\,T}\;
\frac{T^{c}_{1}-T}{T-T^{c}_{2}}  
\end{equation} 
This means that, in the given range of temperatures, the homogeneous
solution $\bar{n}/2$ is \emph{metastable} in an infinite system, because
it is stable only against fluctuations with large wave numbers, i.e.
against small-scale fluctuations. Large-scale fluctuations destroy the
homogeneous state and result in a spatial separation process, i.e.
instead of a homogenous distribution of opinions, individuals with the
same opinion form separated {\em spatial domains} which coexist. The
range of the metastable region is especially determined by the value of
$\alpha<2\mu$, which defines the difference between $T^{c}_{1}$ and
$T^{c}_{2}$.
\item[(iii)] For $ T < T^{c}_{2}$ both eigenvalues $\lambda_1(\bbox{k})$
  and $\lambda_2(\bbox{k})$ are positive for all wave vectors $\bbox{k}$
  (except $\bbox{k} = 0$, for which $\lambda_2=0$ yields), which means that
  the homogeneous solution $\bar{n}/2$ is {\it completely unstable}.  On
  the other hand all systems with spatial dimension $L < 2\pi/k_c$ are
  stable in this temperature region.
\end{enumerate} 
For case (2), \mbox{$T^{c}_{1} < T^{c}_{2}$}, which corresponds to
$\alpha > 2\mu$, already small inhomogeneous fluctuations result in an
instability of the homogenous state for $T < T_2^c$, i.e we have a direct
transition from the completely stable to the completely unstable regime
at the critical temperature $T = T^{c}_{2}$.
 
That means the second critical temperature $T^{c}_{2}$ marks the
transition into complete instability. The metastable region, which exists
for $\alpha <2\mu$, is bound by the two critical social temperatures,
$T^{c}_{1}$ and $T^{c}_{2}$. This allows us again to draw an analogy to
the theory of phase transitions \cite{ulbr-schm-mah-schw-88}.  It is well
known from phase diagrams that the density-dependent {\it coexistence}
curve $T^{c}_{1}(\bar{n})$ divides stable and metastable regions,
therefore we can name the critical temperature $T^{c}_{1}$,
\eqn{crit-temp}, as the {\it coexistence} temperature, which marks the
transition into the metastable regime. On the other hand, the metastable
region is separated from the completely unstable region by a second curve
$T^{c}_{2}(\bar{n})$, known as the spinodal curve, which defines the
region of {\it spinodal decomposition}. Hence, we can identify the second
critical temperature $T^{c}_{2}$, \eqn{temp-2}, as the {\it instability}
temperature.
 
We note that similar investigations of the critical system behavior can
be performed by discussing the dependence of the stability conditions on
the ``social strength'' $s$ or on the total population number $N=A
\bar{n}$. These investigations allow the calculation of a phase diagram
for the opinion change in the model discussed, where we can derive
critical \emph{population densities} for the spatial opinion separation
within the community.

\section{Conclusions} 
We have discussed a simple model of collective opinion formation, based
on active Brownian particles, which represent the individuals. Every
individual shares one of two opposite opinions and indirectly interacts
with its neighbours due to a communication field, which contains the
information about the spatial distribution of the different opinions.
This two-component field has a certain lifetime, which models memory
effects. Furthermore, it can spread out in the community, which describes the
diffusion of information. This way, every individual locally receives
information about the opinion distribution, which affects its further
actions: (i) the individual can keep or change its current opinion, or
(ii) it can stay or migrate towards regions where its current opinion is
supported. Both actions depend (a) on a social temperature, which
describes the stochastic influences, and (b) on the local strength of the
communication field, which expresses the deterministic influences of the
decision of an individual.
 
For supercritical conditions within the community (e.g. supercritical
population size, or supercritical external pressure, or low temperature
etc.), the non-linear feedback between the individuals and the
communication field, created by themselves, results in a process of
\emph{spatial opinion separation}. In this case, the individuals either
change their opinion to match the conditions in their neighbourhood, or
they keep their opinion, but migrate into regions which support this
opinion.

In this paper, we have studied the critical conditions, which may lead to
this separation process. In the spatially homogeneous case, which holds
either for small communities or for an information exchange with infinite
velocity, the communication field can be described in a mean-field
approximation.

For this case, we derived a critical population size, $N^{c}$
(which is related to a critical social temperature, $T^{c}_{1}$). For
$N<N^{c}$, there is a stable balance where both opinions are
shared by an equal number of individuals. For $N>N^{c}$, however, one of
these opinions becomes preferred, hence, majorities and minorities
appear in the
community. Further, we have shown how these majorities change if
we consider an external support for one of the opinions. We found, that
beyond some critical support, the supported subpopulation must always
exist as a majority, since the possibility of its minority status simply
disappears.

As a second case, we have investigated a spatially inhomogeneous
communication field, which is locally coupled to the distribution of the
individuals. This coupling is due to an adiabatically fast relaxation of
the communication field into a quasistationary equilibrium. 

Using this adiabatic approximation, we were able to derive critical
conditions for a \emph{spatial} separation of opinions. We found that
above the critical population size (or for $T<T^{c}_{1}$), the community
could be described as a metastable system, which expresses stability
against small-scale pertubations.  The region of metastability is bound by a
second critical temperature, $T^{c}_{2}$, which describes the transition
into instability, where every pertubation results in an immediate
separation.  Further, we obtained that the range of metastability is
particularly determined by the parameter $\alpha$, which characterizes
how strong an individual responses to the information received from the
communication field.

Finally, we would like to note that our model of collective opinion 
formation only sketches some basic features of structure formation in 
social systems. There is no doubt, that in real human societies a more 
complex behavior among the individuals occurs, and that decision making 
and opinion formation may depend on numerous influences beyond a 
quantitative description. In this paper, we restricted ourselves to a 
simplified dynamical approach, which purposely stretches some analogies 
between physical and social systems. The results, however, display 
similarities to phenomena observed in social systems and allow an 
interpretation within such a context. So, our model may give rise to 
further investigations in the field of quantitative sociology. 
 
\section*{Acknowledgements} 
The authors would like to thank M.Sc. Krzysztof Kacperski (Warsaw) for
support in preparing the figures and obtaining \eqn{s-crit}. One of us
(JAH) is grateful to Professor Werner Ebeling (Berlin) for his
hospitality during the authors stay in Berlin and to the Alexander von
Humboldt Foundation (Bonn) as well as SFB 555 {\it Komplexe Nichtlineare
  Prozesse} for financial support.
 

\end{document}